\documentclass[a4paper, 12pt]{article}%
\usepackage{epsfig}
\usepackage{booktabs}
\usepackage{amsthm}
\usepackage{hyperref}
\usepackage{amsmath}
\usepackage{amsfonts}
\usepackage{amssymb}
\usepackage{graphicx}
\usepackage{float}%
\providecommand{\U}[1]{\protect\rule{.1in}{.1in}}

\begin{document}

\title{Critical Analysis of the Binomial-Tree approach to Convertible Bonds in the
framework of Tsiveriotis-Fernandes model}
\author{K. Milanov and O. Kounchev}
\maketitle

\begin{abstract}
In the present paper we show that the Binomial-tree approach for pricing,
hedging, and risk assessment of Convertible bonds in the framework of the
Tsiveriotis-Fernandes model has serious drawbacks.


\end{abstract}

\section{Introduction}

In the present research, we address a very important and unanswered so far
question regarding the Binomial-tree approach to the Tsiveriotis-Fernandes
(TF) model for pricing Convertible Bonds (CBs). Namely, does the Binomial-tree
framework provide accurate pricing, hedging and risk assessment? We show on a
set of representative examples that by applying the Binomial-tree methodology
one is unable to provide a consistent analysis of the pricing, hedging and
risk assessment.

An important feature of the pricing of CBs is that similar to the American
options there is no closed form solution, and the numerical computation of the
solution is a challenge due to the free boundaries arising. Respectively, in
our study we will employ the natural properties of CBs which are usually
exploited in practice. Depending on the underlying stock we examine the
profile of CB's price, of CB's sensitivities, Convertible Arbitrage strategy,
and the Monte Carlo $\operatorname*{VaR}$ estimation.

Convertible bonds are a widely used type of contract, playing a major role in
the financing of the companies (\cite{Wilmott}, \cite{Gushchin},
\cite{Spiegeleer}). From a pricing and hedging perspective they are highly
complex instruments. They have the early exercise feature of American options
but in three forms: the option to be converted, the option to be called and
the option to be put. Hence, sometimes they behave like a bond and sometimes
like a stock.

Convertible bonds (or simply "convertibles") are bonds issued by a company
where the holder has the option to exchange (to convert) the bonds for the
company's stock at certain times in the future (\cite{Hull}). The "conversion
ratio" is the number of shares of stock obtained for one bond (this can be a
function of time). If the conversion option is executed, then the rights to
future coupons are lost. The bonds are almost always callable (i.e., the
issuer has the right to buy them back at certain times at predetermined
prices). The holder always has the right to convert the bond once it has been
called. The call feature is therefore usually a way of forcing conversion
earlier than the holder would otherwise choose. Sometimes the holder's "call
option" is conditional on the price of the company's stock being above a
certain level. Some convertible bonds incorporate a put feature. This right
permits the holder of the bond to return it to the issuing company for a
predetermined amount.

Throughout the years different convertible bond pricing methodologies were
developed. The main development was in the area of modeling the CB's price
dynamics, as well as towards design of numerical methods for evaluating the
convertible bond pricing function. The most advanced and popular idea for
modeling CB's price dynamics was introduced in the seminal paper of
Tsiveriotis and Fernandes (\cite{TF}, \cite{Zabolotnyuk}). They have proposed
to split the convertible bond value into two components: a cash-only part
which is subject to credit risk, and an equity part, which is independent of
the credit risk. This leads to a pair of coupled partial differential
equations under certain constraints (in fact boundary and free boundary
conditions) that can be solved to value the price of the convertibles. From
numerical point of view Tsiveriotis and Fernandes have proposed explicit
finite difference method for solving their system of equations. On the other
hand, Hull (\cite{Hull}) has proposed to use Binomial-tree approach for
solving the same system. More precisely, the Hull approach is based on Cox,
Ross and Rubinstein (CRR) tree.

Currently, there are two basic approaches for CB pricing, hedging and risk
assessment. The first one that is based on trees (binomial and trinomial)
(\cite{Hull}, \cite{Gushchin}, \cite{Spiegeleer}, \cite{Citigroup}), and the
second one which is based on finite difference techniques (\cite{TF},
\cite{Ayache}, \cite{Andersen}).

There is a gap in the above studies as they do not provide a complete report
on the methodology performance. By the present paper we want to indicate
essential drawbacks of the Binomial-tree methodology and mistakes that are
made when this methodology is used, in major practice areas as hedging and
risk assessment.

The paper is organized as follows: In section \ref{SectionBinTV} we explain
the Binomial-tree scheme for approximation of the TF model. Our main results
are in section \ref{SectionPerfEv} where we provide the performance valuation.
Finally, in the Appendix in section \ref{SectionTF} we provide a short but
closed and informative outline of the model of Tsiveriotis-Fernandes.

\section{Binomial-tree approximation of the TF model\label{SectionBinTV}}

We follow the Binomial-tree approximation to the TF model that is widely used
in practice (cf. \cite{Hull}). It involves modeling the issuer's stock price.
It is assumed that the stock price process follows geometric Brownian motion
and its dynamics is represented by the usual Binomial-tree of Cox, Ross and Rubinstein.

The life of the tree denoted by $T$ is set equal to the life of the
convertible bond denoted also by $T.$ The value of the convertible bond at the
final nodes (at time $T$ ) of the tree is calculated based on the conversion
option that the holder has at that time $T.$ We then roll back through the tree.

At nodes where the terms of the instrument allow calling back the bond, we
test whether the position of the issuer can be improved by calling the bond.
We also test whether the terms of the instrument allow improvement of the
holder's position by selling back the bond to the issuer. Finally, we test
whether conversion is optimal. This is equivalent to setting the convertible
bond value denoted as usually by $V$ at a node equal to
\[
\max[\min(Q_{1},Q_{2}),Q_{3},Q_{4}];
\]
here $Q_{1}$ is the value given by the rollback (assuming that the bond is
neither converted nor called, nor putted at the node), $Q_{2}$ refers to the
\emph{dirty call price}, $Q_{3}$ refers to the \emph{dirty put price}, and
$Q_{4}$ is the value if conversion takes place.

Following the idea of Tsiveriotis and Fernandes, the value of the bond at each
node is represented as a sum of two components, $V=E+B,$ namely, a component
$E$ that arises from situations where the bond ends up as equity, and a
component $B$ that arises from the situations where the bond ends up as a
debt. In addition, the computation of the equity component $E$ of $Q_{1}$ is
based on \emph{risk-free discount rate}, and the debt component $B$ of the
$Q_{1}$ is based on \emph{risky discount rate}.

In order to complete the credit risk concept of TF model, we assume a non-zero
value for the debt component $B$ only when either cash redemption at maturity
or put back of the bond takes place (\cite{TF} eq. 6 \& 12, \cite{Ayache} eq. 47).

\section{Performance Valuation of the Binomial-tree approximation
\label{SectionPerfEv}}

In our practice we have met a lot of examples for which the Binomial-tree
approach of TF model has unsatisfactory performance. To demonstrate this we
will choose the widely known and typical example presented in the paper of
Tsiveriotis and Fernandes (\cite{TF}, Exhibit 5). Namely, our current
performance evaluations are based on terms and conditions that are given in
Table \ref{tab:1}. \begin{table}[h]
\caption{Example Terms and Conditions}%
\label{tab:1}
\centering
\begin{tabular}
[c]{ll}%
\toprule Parameter & Value\\
\midrule Issue Date & 2-Jan-2002\\
Maturity Date & 2-Jan-2007\\
Conversion & 2-Jan-2002 to 2-Jan-2007 into 1 share\\
Call & 2-Jan-2004 to 2-Jan-2007 at 110\\
Nominal & 100\\
Coupon Rate & 4\% paid semi-annually\\
Day Count Convention & $act/365 $\\
Business Day Convention & $Actual $\\
Risk-Free Interest Rate & 5\% (continuously compounded)\\
Credit Spread & 2\% (continuously compounded)\\
Stock Volatility & 30\%\\
\bottomrule &
\end{tabular}
\end{table}

In the next subsections we exhibit the following inconsistencies to market
expectations about:\newline$\bullet$ profile of the price, delta and gamma
sensitivities;\newline$\bullet$ performance of delta-hedge strategy;\newline%
$\bullet$ movement of probability mass of simulations for, e.g., one day
holding period.

Taking into account these inconsistencies we believe a methodology that is
based on Binomial-tree approach would lead very often to impossibility to make
any consistent analysis.

\subsection{CB Price performance valuation}

Regarding the underlying stock, the CB price has such important properties as
strong-monotonicity and convexity. In this section we show that the CB price
$V$ obtained by means of the Binary tree method, misses the strong
monotonicity and convexity, and exposes spurious oscillations. This
misbehavior is persistent no matter how many steps of the Binomial-tree method
we use.

\begin{figure}
[H]
\begin{center}
\includegraphics[
height=2.9563in,
width=3.9243in
]%
{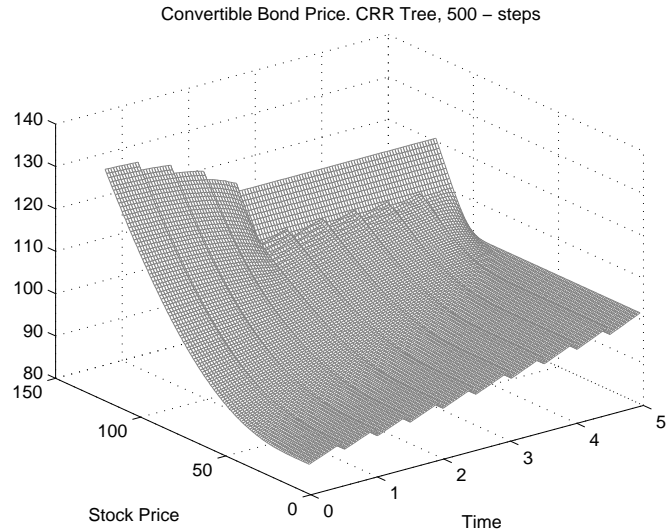}%
\caption{{}CB price, 500 steps}%
\label{fig1}%
\end{center}
\end{figure}

\begin{figure}
[H]
\begin{center}
\includegraphics[
height=2.9563in,
width=3.9287in
]%
{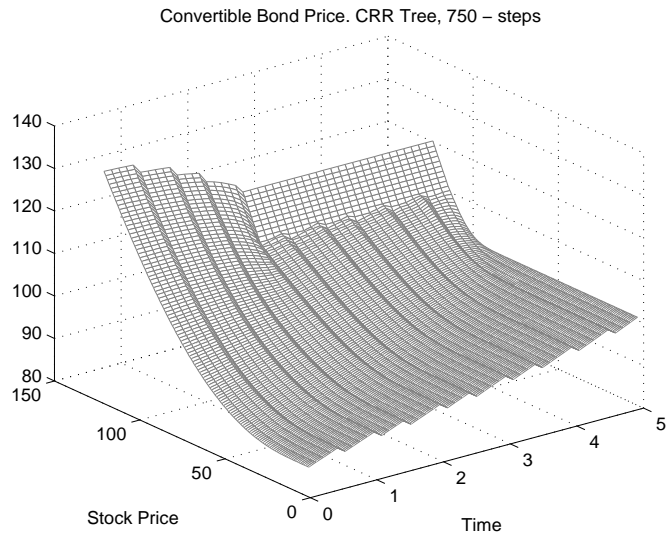}%
\caption{{}CB price, 750 steps}%
\label{fig2}%
\end{center}
\end{figure}

\begin{figure}
[H]
\begin{center}
\includegraphics[
height=2.9563in,
width=3.9243in
]%
{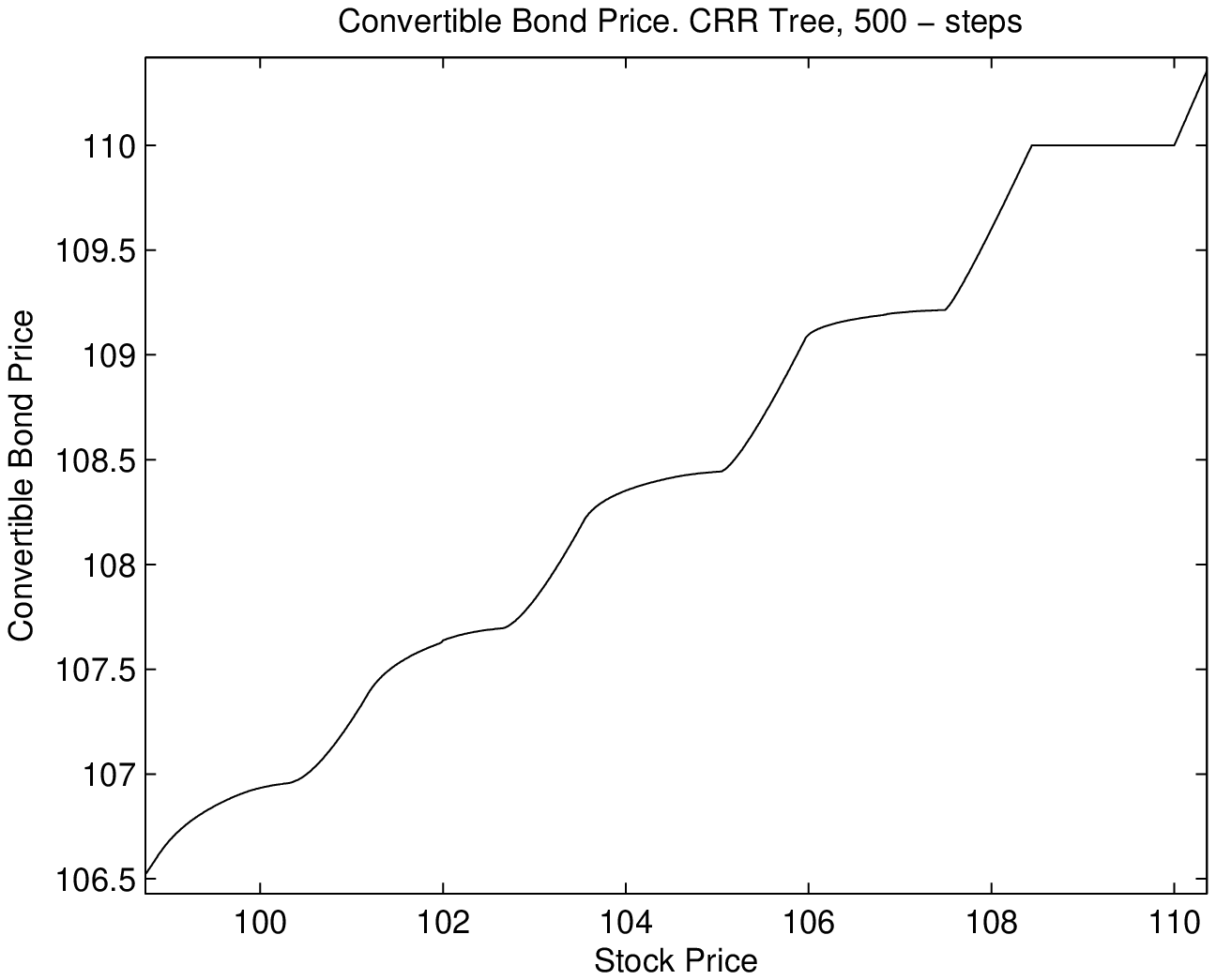}%
\caption{CB price profile at $t=2$}%
\label{fig3}%
\end{center}
\end{figure}

\begin{figure}
[H]
\begin{center}
\includegraphics[
height=2.9563in,
width=3.9776in
]%
{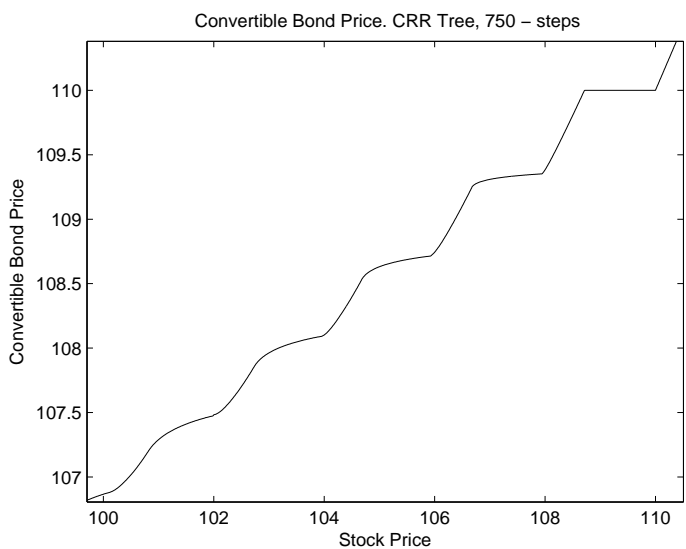}%
\caption{{}CB price profile at $t=2$}%
\label{fig4}%
\end{center}
\end{figure}

To demonstrate the above statements, in Figures \ref{fig1} and \ref{fig2} we
present the evolution of the CB price $V$ during the time till maturity using
a binary tree with 500 and 750 time steps, respectively. The time value 0,
corresponds to the issue date, and the time value 5, corresponds to the
maturity date. For reader's convenience on Figure \ref{fig3} we provide the
section $V\left(  2,S\right)  $ of the CB price $V$ of Figure \ref{fig1}$,$
i.e. we look at $2$ years after the issue date. It is clearly seen that it is
not convex and not strictly monotone in the range of $S$ between $108$ and
$110.$ Also, on Figure \ref{fig4} we provided $V\left(  2,S\right)  $ of the
CB price of Figure \ref{fig2} and have the same observations as above. Let us
emphasize that for $750$ tree steps we have a tree levels at every $1\frac
{1}{2}$ days or so, and for $500$ tree steps -- every $2$ days or so. We make
the final conclusion that even such a detailed Binary tree approximation does
not guarantee a satisfactory result. \textbf{Both figures in identical way
highlight the wrong performance of the approach. }

%

%

%

%

\subsection{CB Delta and Gamma Sensitivities}

\begin{figure}
[H]
\begin{center}
\includegraphics[
height=2.9563in,
width=3.9243in
]%
{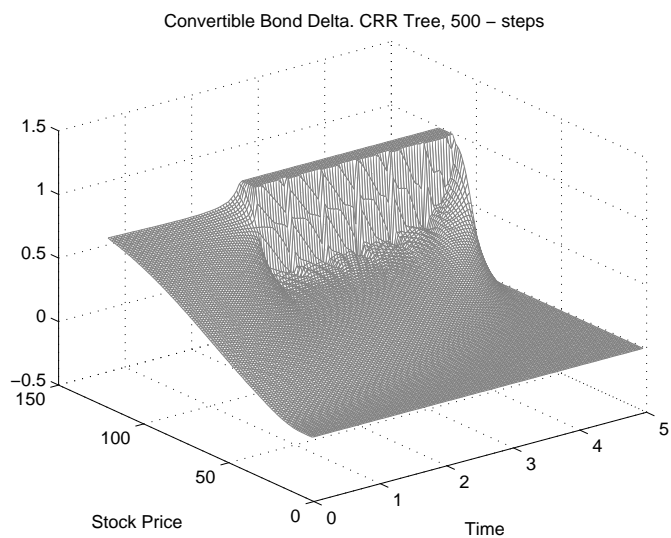}%
\caption{CB delta, 500 steps}%
\label{fig5}%
\end{center}
\end{figure}

\begin{figure}
[H]
\begin{center}
\includegraphics[
height=2.9563in,
width=3.9243in
]%
{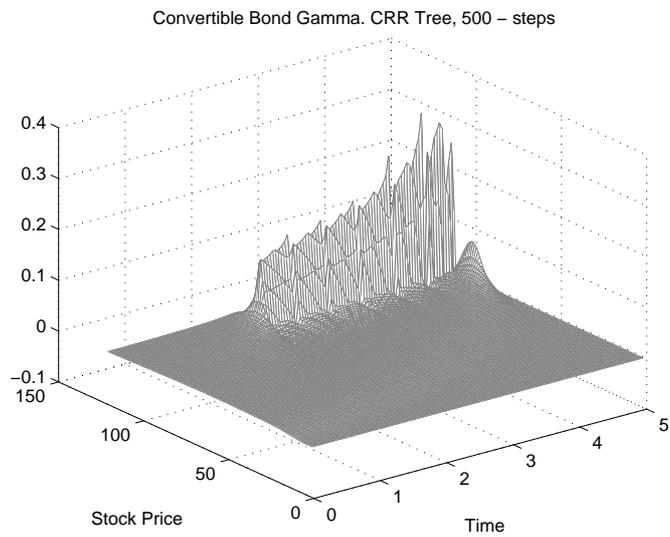}%
\caption{CB gamma, 500 steps}%
\label{fig6}%
\end{center}
\end{figure}

\begin{figure}
[H]
\begin{center}
\includegraphics[
height=2.9563in,
width=3.9243in
]%
{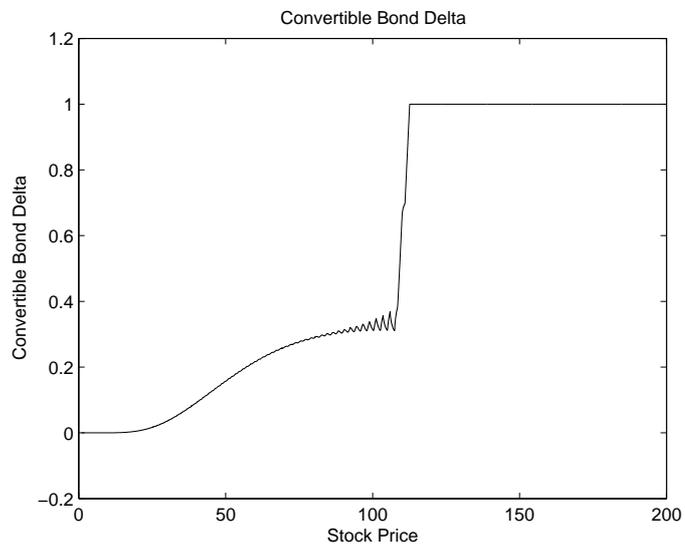}%
\caption{CB delta profile at $t=2$}%
\label{fig7}%
\end{center}
\end{figure}

\begin{figure}
[H]
\begin{center}
\includegraphics[
height=2.9563in,
width=3.9243in
]%
{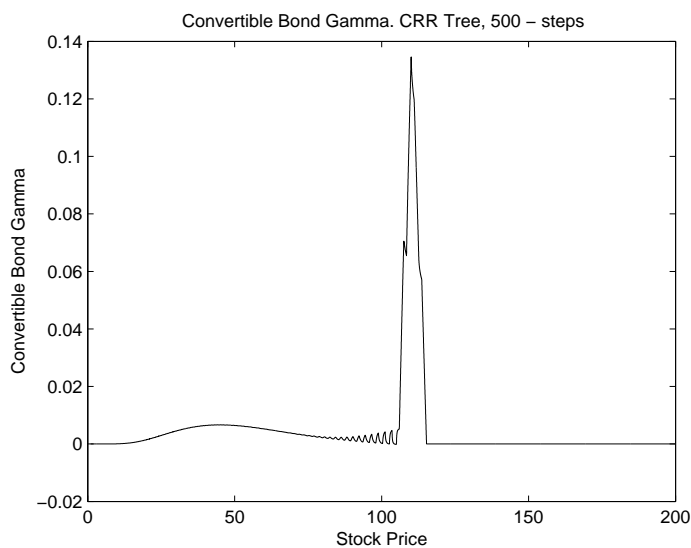}%
\caption{CB gamma profile at $t=2$}%
\label{fig8}%
\end{center}
\end{figure}

Convertible bond delta and gamma quantify the sensitivity of the convertible
price with respect to a small change in the underlying stock. \\*CB delta
sometimes referred to as hedge ratio, is the number of units of the stock we
should short for each CB that we hold in order to create a risk-less
portfolio. On the other hand, it is the slope of the curve that relates the CB
price to the underlying stock price. Thus, the natural definition of CB delta
is
\[
\Delta=\frac{\partial V}{\partial{S}}.
\]
Traders and market makers prefer to use the following form of delta to
illustrate the equity sensitivity of the convertible bond (\cite{Spiegeleer},
p. $23$)
\[
\Delta_{\%}=\frac{\Delta}{C_{r}}%
\]
where $C_{r}$ refers to the conversion ratio (the number of shares per 100
nominal that bond holder gets when converting the bond). This number ranges
between 0 and 100\% whereas the previous delta definition would have values in
the interval $[0,C_{r}]$.\\*CB gamma is a representative measure for convexity
or non-linearity of the instrument. It measures the change in $\Delta$ for a
change in the price of the underlying common stock.
\[
\Gamma=\frac{\partial^{2}V}{\partial^{2}S}=\frac{\partial\Delta}{\partial S}%
\]
From a hedging point of view, CB gamma illustrates how often the position must
be re-hedged in order to maintain a delta-neutral position. That is, if gamma
is small, delta changes slowly, and adjustments to keep a position
delta-neutral need to be made only relatively infrequently. However, if the
absolute value of gamma is large, delta is highly sensitive to the price of
the underlying asset. It is then quite risky to leave a delta-neutral position
unchanged for any length of time.\\*As we have seen, the path dependency and
the possibilities of terminating the bond before the maturity date, prohibit
the derivation of a closed form pricing formula. Thus, the absence of closed
form formula imposes the use of numerical methods to calculate the
Greeks.\\*Finally, let us remark that it is a notorious fact that finite
differences provide a bad approximation to delta and gamma, and are also
computationally expensive. A satisfactory approach has been given for the
computation of delta in (\cite{Hull}, \cite{Wilmott}) and for the computation
of gamma, in (\cite{Wilmott}), and we will follow these references. \\*Within
the Binomial-tree framework convertible bond delta is defined by (cf.
\cite{Hull}, p. $398,$ formula (1.8), p. $170$ ):
\[
\Delta=\Delta\left(  t,S\right)  =\frac{V^{+}-V^{-}}{(u-d)S},
\]
where $t$ is the time and $S$ is the stock price at time zero; $u$ and $d$ are
the parameters of the CRR tree, and $V_{1,1}$ and $V_{1,0}$ are estimated
convertible bond values at one step forward when the stock price is $uS$ and
$dS$ respectively.\\*In a similar way, the convertible bond gamma is defined
by
\[
\Gamma=\Gamma\left(  t,S\right)  =\frac{\Delta^{+}-\Delta^{-}}{(u-d)S},
\]
where $\Delta^{+}$ is the value of delta at one step forward for the stock
price $Su$, and $\Delta^{-}$ is the value of delta at one step forward for the
stock price $Sd$.\footnote{We have to note that although the expressions for
$\Gamma$ are different in \cite{Hull} and \cite{Wilmott}, they provide the
same approximation.}\\*Now, let us come back to market activities as pricing
(Dollar Nuking or Delta Neutral pricing, \cite{Spiegeleer}), analyzing and
hedging where the existence of delta is of crucial importance.\\*Using the
example from Table \ref{tab:1}, in this section we demonstrate that throughout
half of the CB life-span there exist stock prices for which the convertible
bond delta is not well defined and atypically oscillates although the
computation that we have made where based on large number of time steps.
Similarly to the results of delta, the results for the convertible bond gamma
are quite inconsistent.\\*In support of the above statements, in Figure
\ref{fig5} and Figure \ref{fig6} we present the evolution of the CB delta and
gamma during the time till maturity using a binary tree with 500 time steps.
As before, the time value 0 corresponds to the issue date, and the time value
5, corresponds to the maturity date. Also, in order to ease the reader, we
look at 2 years after the issue date, and in Figure \ref{fig7} and Figure
\ref{fig8} we exhibit the profile of CB delta and gamma on the basis of 500
tree steps, namely the one-dimensional sections $\Delta\left(  2,S\right)  $
and $\Gamma\left(  2,S\right)  .$ Both figures in identical way highlight the
wrong performance of the approach corresponding to convertible bond delta and gamma.%

%

%

%

\subsection{Delta-Hedging (Convertible Arbitrage) of CB}

The Delta-hedging in the case of CBs is called Convertible arbitrage.

Convertible arbitrage is a market-neutral investment strategy often employed
by hedge funds (arbitrageurs). It involves the simultaneous purchase of
convertible securities and the short sale of the same issuer's common stock.

The number of shares sold short usually reflects a delta-neutral or
market-neutral ratio. As a result, under normal market conditions, the
arbitrageur expects the combined position to be insensitive to fluctuations in
the price of the underlying stock.

A main reason for the popularity of Binary tree methods is that in the
financial math community the following myths are widely spread: first, the
delta-hedging is only possible in Binary tree framework and Black-Scholes
framework, and second, the Binomial delta becomes, in the limit of time, the
BS delta.

In real life situations the arbitrageurs expect that the Hedged position is
insensitive with respect to the fluctuations in the price of the underlying stock.

In the following example we provide the graph of the relative change of the
convertible arbitrage strategy (hedged position) calculated by means of the
Binomial-tree, where the shock of the stock price is equal to $0.5$. The
contract size of the position of CB given by Table \ref{tab:1} is $1.000.000,$
which is a realistic example. We assume that the settlement date is the issue
date $t.$ The Delta-Hedged position (representing the convertible arbitrage
strategy) is given by
\[
\Pi\left(  S\right)  =V\left(  S,t\right)  -\frac{\partial V\left(
S,t\right)  }{\partial S}\cdot S
\]
while its variation (resulting by the $0.5$ shock) is given by
\[
\Pi\left(  S+0.5\right)  =V\left(  S+0.5,t\right)  -\frac{\partial V\left(
S,t\right)  }{\partial S}\cdot\left(  S+0.5\right)  .
\]
The \textbf{increment }(the change) $\Pi\left(  S\right)  $ of the portfolio
is given by the difference
\[
\Pi\left(  S+0.5\right)  -\Pi\left(  S\right)  =V\left(  S+0.5,t\right)
-V\left(  S,t\right)  -0.5\cdot\frac{\partial V\left(  S,t\right)  }{\partial
S}.
\]

\begin{figure}
[H]
\begin{center}
\includegraphics[
height=2.9563in,
width=3.9243in
]%
{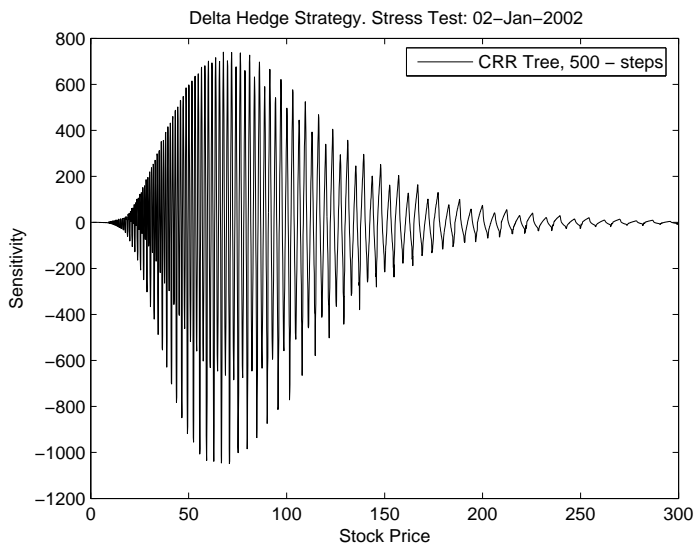}%
\caption{Delta Hedge strategy Stress Test}%
\label{fig9}%
\end{center}
\end{figure}

On Figure \ref{fig9} below we see that the Binomial-tree with $500$ steps does
not meet the expectations of the arbitrageur since it oscillates considerably.

%

\subsection{Risk Assessment}

As an example of the bad performance of the Binomial-tree approximation to
Risk Assessment we will present a simple case of Market Risk Assessment.\\*%
Market risk assessment explores the impact of market observable variables over
the value of an investment (single position or a portfolio). Such variables
are stock prices, interest rates, exchange rates etc. which are sometimes
referred to as market risk drivers or simply risk drivers.\\*A commonly used
methodology for estimation of market risk is \emph{Value-at-Risk} (VaR), (see
e.g. \cite{ZaryFabozzi}). The importance of VaR arises from the fact that
regulators and the financial industry advisory committees recommend VaR as way
of measuring risk. The real boost in the use of VaR came when the Basel
Committee on Banking Supervision adopted banks to use VaR as an internal model
to set their capital requirements.\\*The $\operatorname*{VaR}$ measure is the
highest possible loss over a certain period of time $h$ at a given confidence
level (\cite{ZaryFabozzi}, \cite{Spiegeleer}). The $\operatorname*{VaR}$ with
confidence level $1-\alpha$ is defined as
\[
\operatorname*{VaR}\nolimits_{1-\alpha}=-\min\left\{  x{:P(X<=x)>\alpha
}\right\}  .
\]
Here $1-\alpha$ is the confidence level, $\alpha$ usually takes values like
$1\%$ or $5\%$ and $X$ is the change in the portfolio value, i.e.
$X=V_{h}-V_{0}$. As usually portfolio values $V_{0}$ and $V_{h}$ correspond to
the initial time and the end of the holding period. The mostly used holding
period, over which the expected convertible bond loss is calculated, is one
day or one month (22 business days). \\*It is clear that to calculate
$\operatorname*{VaR}$ values we need the probability density function of the
portfolio value. The $\operatorname*{VaR}$ methodologies mainly differ in ways
of constructing the probability density function. The widely used in practice
are the following methodologies (\cite{ZaryFabozzi}, \cite{Spiegeleer}):

\begin{itemize}
\item Parametric method;

\item Historical simulation;

\item Monte Carlo simulation.
\end{itemize}

We will apply the mostly used third method, namely, the MC simulation. The
reason is that the Parametric method is based on delta and gamma valuation and
we have seen in the previous section that their computation by means of the
Binomial-tree is inefficient. Also, the application of the Historical method
would require to tie down the evidence with a partial historical
environment.\\*

Let us point out, that the use of Binomial-tree approach in building a
$\operatorname*{VaR}$ methodology is too inadequate, due to the fact that the
probability density function in many examples of CBs is inadequate. In support
of this statement we consider the typical CB example considered in Table
\ref{tab:1} and provide its simplified $\operatorname*{VaR}$ analysis as
described in Table \ref{tab:2}.

\begin{table}[h]
\caption{VaR analysis conditions}%
\label{tab:2}%
\centering%
\begin{tabular}
[c]{ll}%
\toprule Parameter & Value\\
\midrule Pricing Model & CRR tree with 500 time steps\\
Evaluation Date & 2-Jan-2004\\
Holding Period & 1 day\\
Confidence level & 99\%\\
Source of risk & Underlying Stock, only\\
Stock Price Scenario type & Log-normal: mean = 0.05, variance = 30\%\\
Number of Scenarios & 10000\\
Stock Spot Price & 100\\
\bottomrule &
\end{tabular}
\end{table}

\begin{figure}
[H]
\begin{center}
\includegraphics[
height=2.9563in,
width=3.9776in
]%
{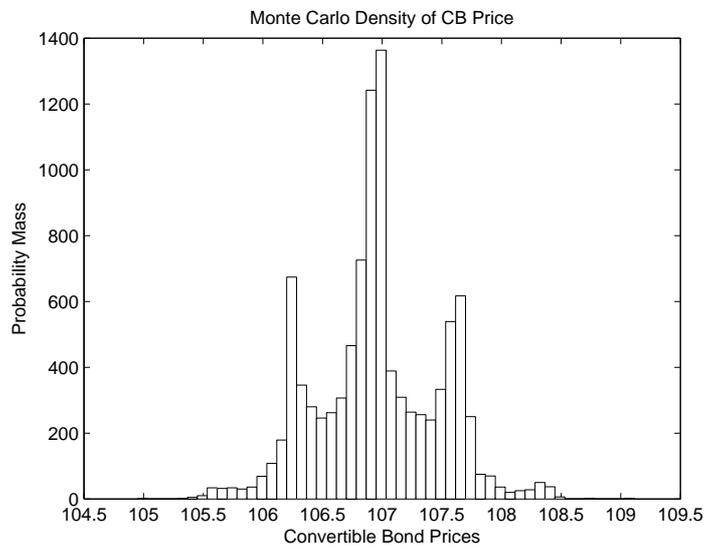}%
\caption{MC density of CB price}%
\label{fig20}%
\end{center}
\end{figure}

\begin{figure}
[H]
\begin{center}
\includegraphics[
height=2.9563in,
width=3.9776in
]%
{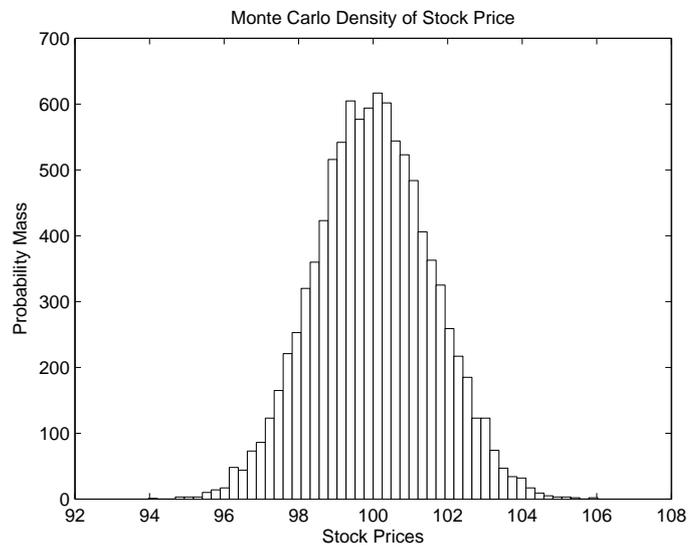}%
\caption{MC density of stock price}%
\label{fig21}%
\end{center}
\end{figure}

Although the $\operatorname*{VaR}$ value at $99\%$ given by
$\operatorname*{VaR}_{99}=1.1285\%$ looks good as a level of risk, in fact it
and all other $\operatorname*{VaR}$ values are very questionable since they
are obtained from the wrongly constructed probability density $\rho.$ Indeed,
we compute the density $\rho$ by means of MC simulation for the underlying
stock that are valid for the end of the holding period. In Figure \ref{fig20}
we see that $\rho$ exhibits the atypical movements of probability mass caused
by the atypical price profile of CB price at the end of the holding period.
Finally, for completeness sake, in Figure \ref{fig21} we show Monte Carlo
scenarios for the underlying stock which are valid for the end of the holding date.%

%

\section{Conclusion}

In this paper we have made performance evaluation of the widely used and
popular techniques of Binomial-tree for approximation of the
Tsiveriotis-Fernandes model for price dynamics of CBs. Our results show that
in many typical examples the Binomial-tree techniques do not meet
practitioners' criteria. Let us mention that even the simplest FDS technique
(the explicit method) has much better performance and this will be the subject
of our next paper.

\section{Appendix. The model of Tsiveriotis-Fernandes (TF) \label{SectionTF}}

For reader's convenience in the present section we provide the
Tsiveriotis-Fernandes (TF) model for computation of the Convertible Bonds.

The pricing of CB has two main periods: before $1998$, and after $1998$ when
the Tsiveriotis-Fernandes model has appeared, \cite{TF}. It represents a major
breakthrough in the area which revolutionized the price computation.

First of all, the system of TF represents a prettily complicated system of
equations which has solutions with free boundary. This makes it much more
complicated for analysis and numerical solution than the American options. For
that reason a Binomial-tree model represents a very intuitive approximation to
the model of TF, and this is completely analogous to the situation in Options
theory where Binomial-tree models are very popular.

The idea of the TF model is that the CB price $V$ is represented as a sum of
two components%
\[
V=B+E
\]
where $B$ is the \textbf{Cash Component}, and $E$ is the \textbf{Equity
Component}.

$B$ is related to the \textbf{future payments} in \textbf{cash}, given at
moment $t$. Then we can construct a risk neutral portfolio
\[
\Pi=B-\Delta S.
\]
In case of \textbf{no default} in the time interval $[t,t+\delta t]$, for
\[
d\Pi=\Pi_{t+\delta t}-\Pi_{t}%
\]
we have
\[
d\Pi=(B_{t}+\frac{1}{2}\sigma^{2}S^{2}B_{S}S)dt.
\]
On the other hand, \textbf{on default} in the same time interval the model
assumption is that the bond holder will lose all future cash flows, that is
\[
d\Pi=-B.
\]
Because of this, the expected value of $d\Pi$ is equal to
\[
E(d\Pi)=(B_{t}+\frac{1}{2}\sigma^{2}S^{2}B_{S}S-r_{c}B)dt
\]
Finally, from \textbf{non-arbitrage arguments}
\[
B_{t}+\frac{1}{2}\sigma^{2}S^{2}B_{SS}+rSB_{S}-(r+r_{c})B=0.
\]

On the other hand the value $E=V-B$ represents the value of the CB related to
\textbf{payments in equity}, and it should therefore satisfy the
\textbf{Black-Scholes }equation
\[
(V-B)_{t}+\frac{1}{2}\sigma^{2}S^{2}(V-B)_{SS}+rS(V-B)_{S}-r(V-B)=0.
\]
Now, replacing the equation for $B$ we obtain equation
\[
V_{t}+\frac{1}{2}\sigma^{2}S^{2}V_{SS}+rSV_{S}-rV-r_{c}B=0.
\]
Thus we have the system of two equations:%

\[%
\begin{split}
&  V_{t}+\frac{1}{2}\sigma^{2}S^{2}V_{SS}+rSV_{S}-rV-Br_{c}=0\\
&  B_{t}+\frac{1}{2}\sigma^{2}S^{2}B_{SS}+rSB_{S}-(r+r_{c})B=0
\end{split}
\]
\noindent where:

$\qquad V$ - \textbf{price of CB}

$\qquad B$ - \textbf{price of cash-only part of the CB}

$\qquad S$ - \textbf{stock price}, $S\in\lbrack0,\infty)$

$\qquad k$ - \textbf{conversion ratio}

$\qquad N$ - \textbf{nominal (par value) of the CB}

\qquad$r$ - \textbf{risk-free rate}

$\qquad r_{c}$ - \textbf{the yield spread, or credit spread}

$\qquad t$ - \textbf{evaluation date,} $t\in\lbrack0,T]$

$\qquad T$ - \textbf{maturity date}

We have the "conversion function":%
\[
cnv\left(  S,t\right)  :=\left\{
\begin{array}
[c]{l}%
k\cdot S\qquad\text{if }t\in\text{\textbf{Conversion periods} of the
Contract}\\
0\qquad\text{otherwise}%
\end{array}
\right.
\]

The "Put Back function" is defined by
\[
B^{Put}\left(  t\right)  =\left\{
\begin{array}
[c]{l}%
b\left(  t\right)  \qquad\text{for }t\in\text{ \textbf{Put periods} of the
Contract}\\
0\qquad\text{otherwise}%
\end{array}
\right.
\]
for a Contracted function $b\left(  t\right)  ;$

The "Call Back function" is defined by
\[
B^{Call}(t)=\left\{
\begin{array}
[c]{l}%
c\left(  t\right)  \qquad\text{for }t\in\text{ \textbf{Call periods} of the
Contract}\\
+\infty\qquad\text{otherwise}%
\end{array}
\right.
\]

We have the following \textbf{Boundary Conditions} and \textbf{Constraints}:%

\[%
\begin{array}
[c]{l}%
\mathbf{Expiry}\text{ }\mathbf{Conditions\quad}\text{for }t=T\\
\quad V(S,T)=\max(cnv(S,T),N)\\
\quad B(S,T)=\left\{
\begin{array}
[c]{ll}%
N, & cnv(S,T)\leq N\\
0, & \text{otherwise}%
\end{array}
\right.
\end{array}
\]
and the
\[%
\begin{array}
[c]{l}%
\mathbf{Boundary}\text{ }\mathbf{Conditions\quad}\text{for }S=0\text{ and
}S=\infty\\
\quad%
\begin{array}
[c]{l}%
\text{when }S=0,\ 0\leq t\leq T\\
\qquad V(0,t)=\max(B^{Put}(t),V(t))\\
\qquad B(0,t)=\max(B^{Put}(t),B(t))\\
\\
\quad\quad\text{where }V=V(t)\text{ and }B=B(t)\text{ are}\\
\qquad V(t)=B(t)=N\cdot e^{-(r+r_{c})(T-t)}\\
\\
\text{when }S\longrightarrow\infty:\ \text{for }0\leq t\leq T\\
\qquad V(S,t)=cnv(S,T),\ \ B(S,t)=0
\end{array}
\end{array}
\]

The \textbf{Payoff Constraints}\textit{:\ }
\[%
\begin{array}
[c]{l}%
V(S,t)=\max(B^{Put}(t),cnv(S,t),\min(B^{Call}(t),V_{held}(S,t)))\\
0\leq S<+\infty,\ 0\leq t<T
\end{array}
\]

We have to note that all conditions and constraints above are for zero coupon
CB which is enough for our present considerations.

E-mail adresses: K. Milanov, kpacu.milanov@gmail.com ; O. Kounchev, kounchev@gmx.de

\end{document}